%% file: Julian_YbRh2Si2.tex
\documentclass[pss]{wiley2sp} 
\usepackage{amsmath}
\usepackage{graphicx}
\usepackage[margin=1in]{geometry}

\begin{document}

\title{
Observation of the J-sheet of the Fermi surface of YbRh$_2$Si$_2$
}

\titlerunning{ $\rm YbRh_2Si_2$ Fermi surface}

\author{%
  A.B.\ Sutton\textsuperscript{\textsf{\bfseries 1}},
  P.M.C.\ Rourke\textsuperscript{\textsf{\bfseries 1,2}},
  V.\ Taufour\textsuperscript{\textsf{\bfseries 3}}, 
  A.\ McCollam\textsuperscript{\textsf{\bfseries 1,4}}, 
  G.\ Lapertot\textsuperscript{\textsf{\bfseries 3}}, 
  G.\ Knebel\textsuperscript{\textsf{\bfseries 3}}, 
  J.\ Flouquet\textsuperscript{\textsf{\bfseries 3}}, 
  S.R.\ Julian\textsuperscript{\Ast,\textsf{\bfseries 1}}  }

\authorrunning{A.\ Sutton et al.}

\mail{e-mail
  \textsf{sjulian@physics.utoronto.ca}, Phone:
  +01-416-978-8188, Fax: +01-416-978-2357}

\institute{%
  \textsuperscript{1}\,Department of Physics, University of Toronto, 
   Toronto, Canada, M5S 1A7\\
  \textsuperscript{2}\,H.H.\ Wills Physics Laboratory, University of 
    Bristol, Bristol, BS8 1TL, United Kingdom\\
  \textsuperscript{3}\,DRFMC, SPSMS, CEA Grenoble, 17 rue des Martyrs, 
    3805-4 Grenoble cedex 9, France\\ 
  \textsuperscript{4}\,High Field Magnet Laboratory, Toernooiveld 7, 
    6525 ED Nijmegen, The Netherlands} 

\received{XXXX, revised XXXX, accepted XXXX} 
\published{XXXX} 

\pacs{ } 

\abstract{%
We report a previously unobserved quantum oscillation frequency 
    in $\rm YbRh_2Si_2$.
This is the first quantum oscillation that can 
    definitely be assigned to the larger of the two major 
    sheets of the Fermi surface predicted by band structure 
    calculations.
Previously observed frequencies were interpreted 
    in terms of a `large' Fermi surface,  
    which includes the Yb 4$f$-hole in the Fermi volume, 
    that has been so strongly spin-split that it 
    closely resembles the `small' Fermi surface,  
    which does not include the Yb 4$f$-hole in the 
    Fermi volume. 
The new frequency can also be incorporated into this picture,  
    however there are some 
    indications that the situation is more complicated 
    than was assumed, 
    suggesting a need for more advanced energy band calculations. 
}

\maketitle   

\section{Introduction}

\begin{figure*}[htb]%
\includegraphics*[width=0.7\textwidth]{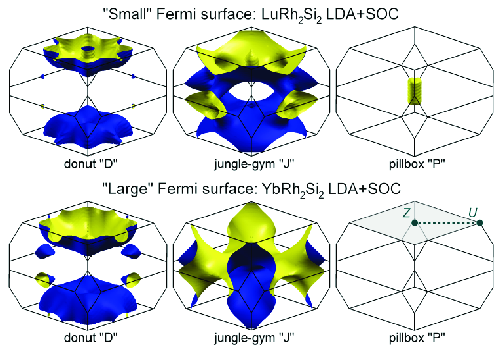}
\caption[]{Top: The calculated `small' Fermi surface, consisting of three sheets. Bottom: the calculated  `large' Fermi surface, consisting of two sheets. The D and J sheets of the Fermi surface appear in both scenarios, and indeed to a good approximation the `large' Fermi surface can be obtained by a rigid shift of the Fermi level of the `small' band structure, and vice-versa \cite{Rourke09a}.
\vskip 3cm
}
\label{fig-FS}
\end{figure*}

The circumstances in which the $f$-electrons are, or are not, included 
      in the volume of the Fermi surace is a central issue in heavy fermion 
      physics in general, and in the theory of quantum criticality of 
      heavy fermions in particular
      \cite{Gegenwart08,Senthil04}.
Generally only the two extreme cases are considered to be permitted: 
      either the Fermi suface is `large,' in which case the f-electrons or f-holes 
      are included in the Fermi volume, or it is `small,' and they are not.
A particularly elegant demonstration of a change of Fermi volume  
      across a quantum critical point was found in CeRhIn$_5$ under 
      pressure \cite{Shishido05}, where de Haas-van Alphen oscillation 
      measurements were used to measure the Fermi surface on either side of a 
      pressure-induced quantum critical point separating antiferromagnetic 
      and paramagnetic ground states.

YbRh$_2$Si$_2$ has played an important role in this discussion.  
Very modest magnetic fields are sufficient to suppress 
      a weak antiferromagnetic state, producing a field-induced 
      quantum critical point \cite{Custers03}
      or possibly a quantum critical phase  \cite{Friedemann09}. 
A central feature of this quantum critical point is argued to be a 
      change of the Fermi volume from small at low field to large at high field. 
The primary evidence for this comes from the Hall coefficient, 
      in which a rapid change tends towards a 
      discontinuous jump in the zero-temperature limit 
      \cite{Paschen04}. 

In principle, Fermi surface measurements can prove that such a 
      Fermi surface transformation takes place. 
However, ARPES, which can only be done in zero field, cannot 
      yet be carried out at sufficiently low temperatures,  
      although it has yielded useful information at higher temperatures 
      \cite{Wigger07}. 
Quantum oscillation measurements such as the 
      de Haas-van Alphen effect (dHvA), on the other hand, 
      can be carried out at millikelvin temperatures, but require high 
      magnetic fields. 
It would seem obvious that dHvA can at least verify that the FS is indeed 
      large at high field,  which would provide partial confirmation of the 
      proposed scenario. 
However application of high magnetic fields to YbRh$_2$Si$_2$ has 
      dramatic effects, producing a remarkable fall in the linear 
      specific heat coefficient $\gamma$ and in the $T^2$ coefficient of 
      resisitivity, suggesting 
      heavy fermion/Kondo physics is rapidly suppressed as the field 
      increases from the quantum critical field up to 10 T; 
      but above 10 T the suppression seems to be 
      frozen part way, with $\gamma$ still around $\rm 100\ mJ/mol\,K^2$ 
      \cite{Tokiwa05,Gegenwart06}. 
Additionally, between 
      the QCP and 10 T the moment per Yb grows from zero to just 
      above $1 \mu_B$, but above 10 T this too becomes comparatively 
      independent of applied field \cite{Tokiwa05}. 
Within the `large' Fermi surface scenario such a magnetization must 
      arise from a substantial spin-splitting of the Fermi surface, 
      whereas in the `small' Fermi surface scenario it would come 
      from polarization of local moments, which would themselves 
      induce a small spin-splitting of the conduction bands such as 
      is seen in the ferromagnetic system CeRu$_2$Ge$_2$ \cite{King91} 
      or gadolinium \cite{Ahuja94}. 
The problem then arises of distinguishing between an extreme, spin-split 
      version of the large  Fermi surface and a spin-polarized version of 
      the small, and this was addressed in our recent 
      paper \cite{Rourke09}. 

Quantum oscillation measurements map the Fermi surface via the 
      Onsager relation, $F = \hbar A/2\pi e$, which connects 
      oscillation frequencies with extremal cross-sectional areas 
      of Fermi surface \cite{Shoenberg}.  
In practice, it is necessary to compare the angle dependence of measured 
      oscillation frequencies with the predictions of calculated 
      band structures.
In the case of YbRh$_2$Si$_2$ there are now a number of such 
     calculations for both the large and small Fermi surfaces 
     \cite{Wigger07,Norman05,Knebel06,Rourke09,Rourke09a}, and the Fermi surfaces 
     shown in figure \ref{fig-FS} are representative. 
Previous work \cite{Knebel06,Rourke09} found dHvA oscillation 
     frequencies whose angle dependence 
     matched neither the predicted large nor small Fermi surface exactly, 
     but seemed closer to small,  
     because some of the observed frequencies could be explained by 
     orbits that pass through the `hole' in the donut, or `D,' surface, 
     as described below.  
The D surface of the large Fermi surface does not have this hole 
     (compare the the upper and lower surfaces on the left in 
     figure \ref{fig-FS}). 
This was not interpreted as evidence that the Fermi surface of 
     YbRh$_2$Si$_2$ is actually small at high field, 
     which would directly contradict the quantum criticality scenario 
     of reference \cite{Paschen04}. 
Rather, an unusual field dependence of the dHvA frequencies below 
     11 tesla led us to propose that the apparent small Fermi surface 
     was in fact a spin-polarized version of the large   
     Fermi surface, which is produced when the magnetic polarization of 
     the large Fermi surface is interrupted by a Lifshitz 
     transition, after which the majority spin Fermi surface is trapped in the 
     hybridization gap.  
Such a scenario has theoretical backing \cite{ViolaKusminsky08,Beach08}, 
     and was originally proposed for the metamagnetic transition of 
     CeRu$_2$Si$_2$ \cite{Daou06}. 

A weakness of the previous dHvA studies is that all of the observed 
     frequencies were assigned to the D sheet, and none to the 
     J-sheet, which is certainly equally important thermodynamically.
This left open the actual interpretation of the oscillation frequencies, 
     and did not rigorously test the accuracy of band structure calculations. 
Here, we report further quantum oscillations studies, 
     including the observation of a new, high frequency that we 
     believe must originate on the J sheet of the Fermi surface.

\section{Experiment}

\begin{figure*}[htb]%
\includegraphics*[width=0.7\textwidth,height=10cm]{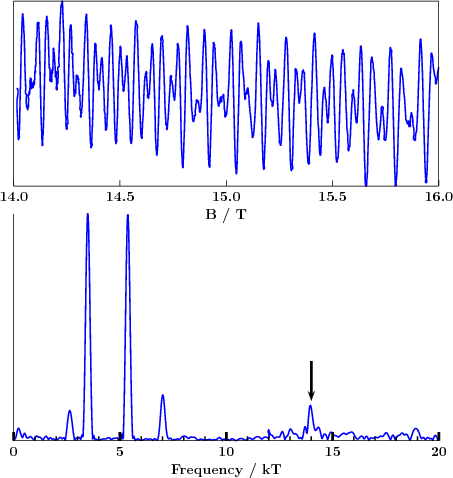}
\caption{Top: Oscillatory magnetization from 14 to 16 T, for 
field along the (100) direction.  Bottom: Fourier transform showing 
quantum ocillation frequencies.  
The frequency spectrum above 12 kT has been multiplied 
by 10 to make the 14 kT peak, indicated by an arrow, more visible.  
\vskip 3cm
}
\label{fig-oscs}
\end{figure*}

De Haas-van Alphen oscillations were measured using the 
    standard field modulation technique \cite{Shoenberg}. 
The samples were mounted on a cryomagnetic system consisting of a 
    dilution refrigerator with a 
    base temperature below 20 mK and a 16/18 T superconducting magnet. 
The samples are platelets, grown from indium flux \cite{Knebel06}, 
    with typical dimensions of 1 mm $\times$ 2 mm $\times$ 0.2 mm, 
    which is somewhat smaller than optimal for these measurements. 
The main difference between the present measurements and 
    those of reference \cite{Rourke09} are a number of improvements 
    in signal-to-noise ratio for these very weak signals, including 
    the careful construction of pick-up coils that give the largest 
    possible filling factor. 
The coils were placed in a rotation mechanism, and the field 
    direction was varied in steps of about 2 degrees.  
In this paper we focus on measurements in the $a$-$b$ plane, 
    where the new high frequency was observed. 

\section{Results}

The upper plot in 
    figure \ref{fig-oscs} shows a quantum oscillation trace between 14 and 
    16 T with the field along the (100) direction, while 
    the lower trace shows the Fourier spectrum of this trace. 
In the Fourier spectrum the data above 12 kT have been multiplied by  
    a factor of 10 to make the new peak, indicated by an arrow, 
    more visible. 
Although the intensity of this signal is weak, it extends over an 
    angular range of about 10 degrees, as seen in 
    the right-hand panel of figure \ref{fig-angledep}(c), which shows the 
    angle dependence of observed frequencies in the $a$-$b$ plane.
It is necessary, in identifying this frequency as a new orbit, to exclude 
    the possibility that it is the second harmonic of the peak that 
    is seen at 7 kT in figure \ref{fig-oscs}. 
For this reason it is important that we have carried out both mass 
    studies, using the Lifshitz-Kosevich formula to obtain the quasiparticle 
    mass from the temperature dependence of the oscillations  \cite{Shoenberg}, 
    and that we have examined the angle dependence of the frequencies. 
The masses are given in table \ref{table-masses}, but 
    they are inconclusive, since the mass of the 14 kT peak is 
    close to twice that of the 7 kT peak, as would be expected of a 
    second harmonic.  
However, it is not unusual for masses to scale with frequency, as was found 
    for example in UPt$_3$ in the first detailed dHvA study of a heavy fermion 
    metal \cite{Taillefer88}, so this does not prove that the 14 kT oscillation
    is a harmonic of the 7 kT oscillation. 
Indeed, the angle dependence shows clearly that the 
    ratio of two for these frequencies is a coincidence, because as we rotate the 
    field away from the (100) direction the angle dependence is very different, 
    and the ratio of the frequencies becomes different from two. 
We might similarly worry that the 7 kT is itself a second harmonic of the 
    very strong 3.5 kT peak, however again the angle dependence does 
    not support this in the sense that the 7 kT frequency extends over a 
    much larger angular range than the 3.5 kT frequency, 
indeed, as discussed below, there are good reasons 
    for expecting the frequencies on the `D' sheet to come in pairs, 
    as is the case with these two frequencies. 

\begin{table}[b]
  \caption{  Frequencies and their corresponding masses on the 
(110) and (100)-axes.  
}
  \begin{tabular}[htbp]{@{}lllll@{}}
    \hline
    \multicolumn{2}{ c }{ (110) } & & \multicolumn{2}{ c }{ (100) } \\
    \hline
    $F\ (kT)$  & $m^*/m_e$  & \hspace{5mm} & $F (kT)$ & $m^*/m_e $\\
    \hline
    2.64  & $12.5 \pm 0.5$ &  &  &  \\
    3.48  & $6.1 \pm 0.4$ &  & 3.22 & $ 7.0 \pm 0.1$ \\
    5.37  & $9.2 \pm 0.2$ &  & 5.65  & $8.9 \pm 0.1$  \\
    7.01  & $12.3 \pm 0.3$ &  & 6.15 & $8.44 \pm 0.1$ \\
          &                &  & 6.54 & $13.2 \pm .2$ \\
    14.0  & $21 \pm 2$     &  &      &               \\
    \hline
  \end{tabular}
  \label{table-masses}
\end{table}

\section{Discussion}

\begin{figure}[htb]%
\includegraphics[width=.4\textwidth]{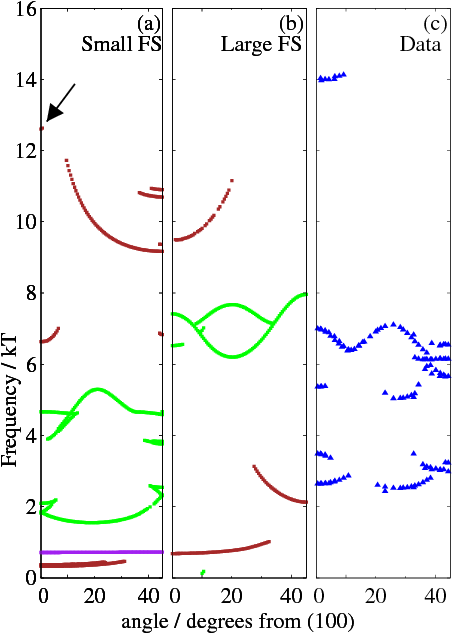}%
\caption{De Haas-van Alphen frequencies vs.\ field angle for fields applied in the basal plane. 
0 degrees corresponds to (100), while 45 degrees is (110).  
Figure (a) is the prediction of the `small' Fermi surface calculation 
(see figure \ref{fig-FS}), (b) is the prediction of the `large' Fermi 
surface calculation, while (c) is the experimentally measured angle dependence. 
In (c), 
the upward curvature of the 14 kT frequency with angle shows that it is not 
the second harmonic of the 7 kT peak, 
which curves down with increasing angle. 
In (a) and (b) the frequencies are colour coded, such that the green 
line corresponds to orbits on the D sheet, brown corresponds to orbits 
on the J sheet, and the lone purple line at low frequency in (a) 
corresponds to the `pillbox' sheet of the Fermi surface.  
The arrow in (a) points to a J orbit that may correspond to the 
14 kT frequency in (c).
}%
\label{fig-angledep}
\end{figure}

In figure \ref{fig-angledep}(c) there are three groups of frequencies: 
    one group between about 2.5 and 3.5 kT, a second between about 5 and 
    7 kT, and the lone high frequency near 14 kT close to the (100) axis. 
The two lowest groups had been observed before \cite{Knebel06,Rourke09},  
    and the remarkable similarity in the angle dependence of frequencies 
    in these two groups was noted, 
    with roughly a factor of two between several of the 
    frequencies, and some frequencies disappearing in one group and 
    reappearing at twice (or one-half) the frequency in the other group. 
The interpretation of this behaviour 
      is that both sets of orbits originate on the D surface, with the lower 
      group arising from orbits that pass through the hole in the donut 
      (see top-left figure in figure \ref{fig-FS}), while those in the upper group 
      encircle the entire Fermi surface. 
This explains the factor of roughly two difference in frequency: 
      they are almost the same orbit, but the lower set only go around 
      half of the surface.  
Of course, as expected, there is not exact duplication, because 
     the orbits that pass through the hole in the donut must be so-called 
     `central' orbits 
     while those that do not go through the hole must be non-central. 
The two groups of frequencies are also seen in the predicted angle 
     dependence of frequencies for the small D surface 
     (green lines in figure \ref{fig-angledep}(a)), but not for the `large' Fermi 
     surface, because the large D Fermi surface does not have a hole. 
This leads to the supposition that the Fermi surface must in this field range be 
     closer in topology to the `small' than to the `large' Fermi surface. 

The new branch at 14 kT, for $B || (100)$ tends to reinforce this interpretation, 
    in the sense that the `small' Fermi surface is predicted to give rise 
    to a high frequency over a very limited range of angles near the (100) 
    axis (indicated by the arrow in figure \ref{fig-angledep}(a)).  
This is closer in frequency than a similar branch on the J surface of the 
    large Fermi surface that starts near 9.5 kT and has a large angular range 
    (see figure \ref{fig-angledep}(b)).  
However, it is surprising, if the small Fermi surface is accurate, 
    that the other predicted orbits from the J surface, 
    extending from about 10 degrees over to the (110) axis, between 
    12 and 9 kT in figure \ref{fig-angledep}(a), have not been 
    observed.  
Although it is generally dangerous to conclude anything 
    from failure to see a quantum oscillation, it seems likely that the 
    required conditions of sample purity and temperature and field are being 
    met. 
In this sense, the J surface of the large Fermi surface is  
    in better agreement with the data, because it does not have any 
    high frequencies near the (110) axis. 

So the interpretation of this new frequency in terms of the large or 
    small Fermi surfaces is ambiguous.
However it is clear that the 14 kT frequency must originate on 
    the J surface regardless of which model is used, as it is much larger 
    than any predicted orbit on the D surface for either scenario. 

We are currently investigating the angle dependence of quantum oscillations 
    in the $a$-$c$ and $b$-$c$ planes, however  
    our results have probably already reached the limit of 
    what can be achieved by 
    comparisons with the LDA calculated `small' and `large' Fermi 
    surfaces. 
One possible approach will be to shift the Fermi levels in these calculations, 
    to look for better agreement with our results, but there is 
    probably a need for more advanced band-structure calculations, 
    either employing dynamical mean-field theory \cite{Shim07}
    or using the renormalized band theory approach of Zwicknagl et al.\ 
    \cite{Fulde88}. 

\section{Conclusions:}

We have observed a new orbit in YbRh$_2$Si$_2$ corresponding to a 
    major sheet of the Fermi surface, and comparison with calculations  
    strongly suggests that this is the first observation 
    of the thermodynamically important `J' sheet of the Fermi surface. 
Interpretation of this orbit in terms of either the `large' or `small' 
    Fermi surface scenario is problematic, and suggests the need for 
    more advanced band structure calculations at high magnetic fields 
    in this material.

\begin{acknowledgement}
This research was funded by the France-Canada Research Fund, 
the Canadian Institute for Advanced Research, and the National 
Science and Engineering Research Council of Canada. 
\end{acknowledgement}

\input{references.tex}

\end{document}

%% file: references.tex
%
%